\documentclass[aps,prl,twocolumn,superscriptaddress]{revtex4}
\usepackage[english]{babel}
\usepackage[utf8]{inputenc}
\usepackage{amsmath}
\usepackage{amssymb}

\usepackage{color,graphicx}
\usepackage{float}
\usepackage{multirow}
\usepackage{hhline}

\begin{document}

\author{Julien Varennes}
\affiliation{Department of Physics and Astronomy, Purdue University, West Lafayette, IN 47907, USA}

\author{Sean Fancher}
\affiliation{Department of Physics and Astronomy, Purdue University, West Lafayette, IN 47907, USA}

\author{Bumsoo Han}
\affiliation{Schools of Mechanical Engineering \& Biomedical Engineering, Purdue University, West Lafayette, IN 47907, USA}

\author{Andrew Mugler}
\email{amugler@purdue.edu}
\affiliation{Department of Physics and Astronomy, Purdue University, West Lafayette, IN 47907, USA}

\title{Emergent versus Individual-based Multicellular Chemotaxis}

\begin{abstract}
    Multicellular chemotaxis can occur via individually chemotaxing cells that are mechanically coupled. Alternatively, it can emerge collectively, from cells chemotaxing differently in a group than they would individually.
    Here we consider collective movement that emerges from cells on the exterior of the collective responding to chemotactic signals, whereas bulk cells remain uninvolved in sensing and directing the collective.
    We find that the precision of this type of emergent chemotaxis is higher than that of individual-based chemotaxis for one-dimensional cell chains and two-dimensional cell sheets, but not three-dimensional cell clusters. We describe the physical origins of these results, discuss their biological implications, and show how they can be tested using common experimental measures such as the chemotactic index.
\end{abstract}

\maketitle

Collective migration is ubiquitous in cell biology, occurring in organism development \cite{theveneau2010collective,cai2016modeling,bianco2007two,montell2008morphogenetic},
tissue morphogenesis \cite{ellison2016cell} and metastatic invasion
\cite{kim2013cooperative,friedl2009collective,friedl2012classifying,deisboeck2009collective}.
Collective migration often occurs in response to chemical cues in the environment, a process known as chemotaxis. The simplest way for cells to collectively chemotax is by individual detection and response to the chemical attractant: each cell measures the spatial difference in chemoattractant across its body and moves in the perceived direction of the gradient, while mechanical coupling keeps the group together. Groups performing this type of \textit{individual-based chemotaxis} (IC) are found throughout cell biology \cite{kulesa1998neural,charteris2014modeling}.
However, recent experiments have uncovered an alternative type of chemotaxis, in which cells grouped together chemotax differently than if they were alone \cite{haeger2015collective,malet2015collective,leber2009molecular,gaggioli2007fibroblast}.
Specifically, outer cells polarize while inner cells do not, a mechanism observed in neural crest cells \cite{theveneau2010collective} and considered in several recent modeling studies
\cite{malet2015collective,camley2016emergent,varennes2016collective}. This type of \textit{emergent chemotaxis} (EC) behavior seen in cell collectives presupposes a machinery within cells which allows for behavior to change once a cell is in a group. Since this machinery may come at a cost, this raises the question of whether EC offers any fundamental advantage over IC.

\begin{figure}
    \centering
        \includegraphics[width=0.45\textwidth]{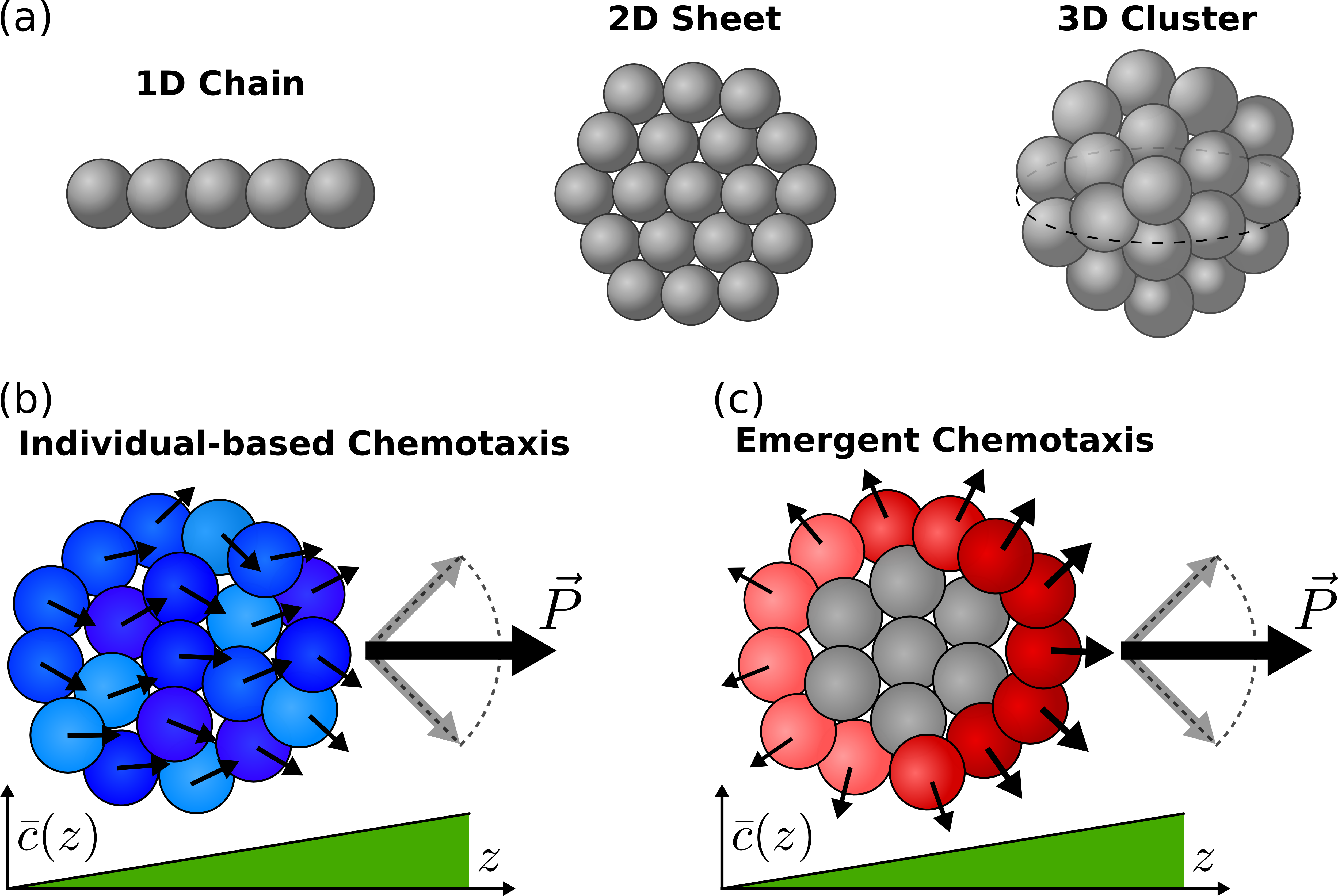}
    \caption{(a) We study the chemotactic performance of 1D chains, 2D sheets, and 3D clusters of cells. (b) In individual-based chemotaxis (IC), cells in the collective polarize based on their own gradient measurement. (c) In emergent chemotaxis (EC), cell polarization depends on intercellular interactions: cells on the edge polarize based on their measurement of the concentration, and cells in the bulk do not polarize. In both mechanisms the total polarization $\vec{P}$ will fluctuate in magnitude and direction due to noise in cell measurements.} \label{fig:1}
\end{figure}

We address this question using simple physical models of EC and IC
\cite{footnote1}.
Cell collectives respond to graded profiles of freely diffusing molecules, and we quantify the migratory behavior of one-dimensional (1D) cell chains, two-dimensional (2D) cell sheets, and three-dimensional (3D) cell clusters [Fig.\ \ref{fig:1}(a)], configurations designed to mimic physiological multicellular structures such as filaments and ducts \cite{cheung2013collective,friedl2009collective,bardeesy2002pancreatic}. Collectives performing EC and IC are found to have very similar mean polarization, but we will show that 1D and 2D EC collectives have higher chemotactic precision than IC collectives: we find that for $N$ cells, the relative error in EC scales as
$\{N^{-2}, N^{-3/2}, N^{-1}\}$ for 1D, 2D, and 3D, respectively, whereas in IC it scales as $N^{-1}$ for any dimension. We explain the physical origin of this difference and discuss its implications.

We first consider IC [Fig.\ \ref{fig:1}(b)]. Due to the chemoattractant molecules in the environment, each cell $i$ becomes polarized with vector $\vec{p}_i$ in its desired direction of motion \cite{jilkine2011comparison}. The components of $\vec{p}_i$ reflect the difference in concentration $c(\vec{r},t)$ between the front and back of the cell in each respective direction.
This concentration difference will fluctuate due to the particulate nature of diffusion \cite{berg1977physics, bialek2005physical}. Focusing on this extrinsic source of noise, we treat each cell as a sphere of radius $a$ through which molecules freely diffuse
\cite{footnote2}, akin to the ``perfect instrument'' described by Berg and Purcell \cite{berg1977physics}.
The concentration difference is encoded internally as a weighted count of the molecules within the cell volume. The weighting function will depend on the sensory network, but will generally be positive at the front and negative at the back; here we choose cosine for simplicity. Orienting our coordinate system such that $\hat{z}$ is parallel to the gradient, the components of $\vec{p}_i$ become
\begin{equation}
    p_{i\alpha}(t) = \int_{U_i} d^3r \ w_\alpha(\hat{r}) \ c(\vec{r},t) , \label{eq:ICcell}
\end{equation}
with $U_i$ the cell volume,
$\alpha\in\{x,y,z\}$, and in spherical coordinates the cosine is
$w_\alpha(\hat{r}) = \{\sin\theta \cos\phi, \sin\theta \sin\phi, \cos\theta\}$.
The concentration $c(\vec{r},t)$ is a random variable
which obeys regular diffusion $\dot{c} = D\nabla^2c+\eta_c$ with $D$ the diffusion coefficient.
The Langevin noise term $\eta_c$ obeys
$\langle \eta_c(\vec{r},t) \eta_c(\vec{r}',t') \rangle = 2 D \delta(t-t') \vec{\nabla}_{r} \cdot \vec{\nabla}_{r'} ( \bar{c}(\vec{r}) \delta^3(\vec{r}-\vec{r}'))$,
and it accounts for the diffusive fluctuations in concentration
\cite{gardiner1985handbook,fancher2016fundamental}.
We first consider a constant gradient $g$ with mean concentration profile
$\bar{c}(\vec{r}) = c_0 + gz$.
Cells are assumed to rigidly adhere to one another, hence the polarization of a collective of $N$ cells is the sum of its constituent cells' polarization vectors $\vec{P}(t) = \sum_{i=1}^N \vec{p}_i(t)$
\cite{footnote3}.

\begin{figure}[t]
    \centering
        \includegraphics[width=0.49\textwidth]{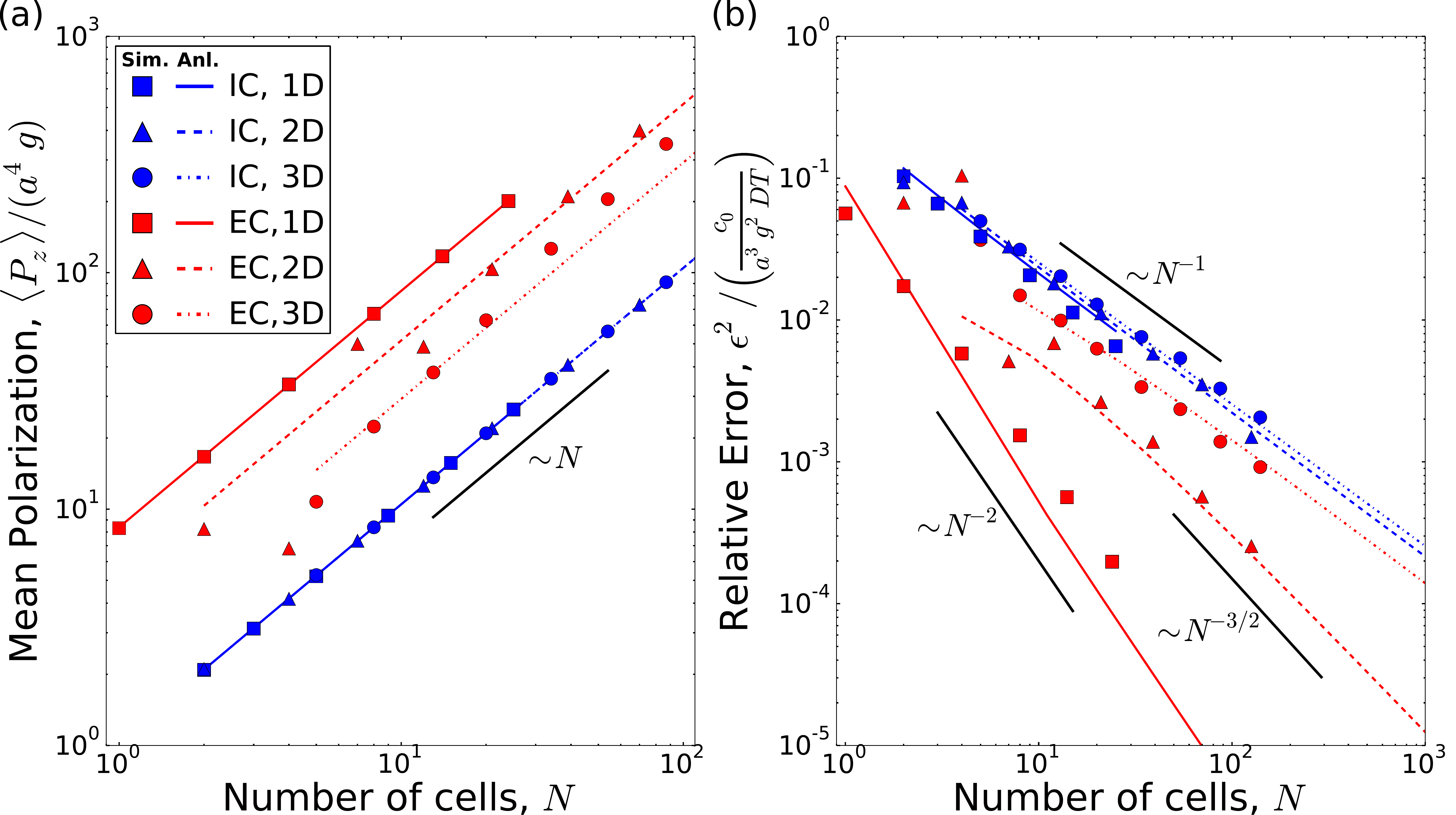}
    \caption{(a) Mean cluster polarization and (b) relative error for both mechanisms of collective chemotaxis in every configuration. Points are simulation data, colored lines are analytical predictions. 1D EC data plotted with respect to $N-1$.} \label{fig:2}
\end{figure}

The collectives exist at low Reynolds number, hence their velocity $\vec{v}$ is proportional to the motility force, and in turn the polarization $\vec{P}$.
Therefore, understanding the behavior of $\vec{P}$ will inform us of the collective migratory performance. We focus on two measures of performance: the mean and the relative error of the polarization in the gradient direction $P_z$, where the relative error is defined
\begin{equation} \label{eq:error1}
    \epsilon^2
    = \frac{\text{Var}[P_z]}{\langle P_z \rangle^2}
    = \frac{\text{Var}[v_z]}{\langle v_z \rangle^2}.
\end{equation}
To investigate $\langle P_z\rangle$ and $\epsilon^2$ for the IC model,
we first perform particle-based simulations of the chemoattractant in the presence of the permeable cells \cite{supinfo,SIcode1}. We find that the total mean polarization $\langle \vec{P} \rangle$ points solely in the gradient direction with equal magnitude regardless of dimensionality [Fig.\ \ref{fig:2}(a), blue data points]. Indeed, Eq.\ \ref{eq:ICcell} indicates that a single cell will have mean polarization proportional to the concentration difference across the cell,
$\langle \vec{p}_i \rangle = \pi a^4g \hat{z}/3$,
regardless of the cell's location.
Therefore the mean collective polarization is geometry-independent, depending only on the number of cells present,
\begin{equation} \label{eq:ICmean}
    \langle \vec{P} \rangle_\text{IC} = \frac{\pi}{3} a^4gN \ \hat{z},
\end{equation}
as shown in Fig.\ \ref{fig:2}(a) (blue lines).

We next investigate the relative error for IC collectives. Simulations show that the error decreases with cluster size as
$\epsilon^2 \sim N^{-1}$
for all three geometries [Fig.\ \ref{fig:2}(b), blue data points]. Indeed this is the result that one would obtain if the cells were independent sensors \cite{simons2004many}. However, they are not independent: their noise is correlated by fluctuations in the concentration \cite{fancher2016fundamental,mugler2016limits}. To understand why correlations do not affect the relative error we investigate the model analytically.

We express the concentration as $c(\vec{r},t) = \bar{c}(\vec{r}) + \delta c(\vec{r},t)$ as well as the cell polarization
$\vec{p}_i(t) = \langle \vec{p}_i \rangle + \delta\vec{p}(t)$,
and by Fourier transforming in both space and time we derive analytic expressions for
$\text{Var}[P_z]$ and thereby $\epsilon^2$ \cite{supinfo}. Since
$P_z = \sum_{i=1}^N p_{iz}$,
the variance in the total polarization is a linear combination of all cell polarization variances and covariances present in the collective,
\begin{equation} \label{eq:ICvar1}
    \text{Var}[P_z] =
    \sum_{i} \text{Var}[p_{iz}]
    + \sum_{i\neq j} \text{Cov}[p_{iz},p_{jz}]
    \equiv V + C ,
\end{equation}
The variance and covariance for cells within the collective are derived from the power spectrum in polarization cross correlations, taking the general form
\begin{align} \label{eq:ICvar2}
\begin{split}
    \text{Cov}[p_{i\alpha},p_{j\alpha}] &= \frac{1}{T}
    \lim_{\omega \to 0} \int \frac{d\omega'}{2\pi}
    \langle\delta\tilde{p}_{i\alpha}^*(\omega')\delta\tilde{p}_{j\alpha}(\omega) \rangle
    \ ,
\end{split}
\end{align}
with $T$ the cell's measurement integration time and $\text{Var}[p_{i\alpha}] = \text{Cov}[p_{i\alpha},p_{i\alpha}]$ \cite{fancher2016fundamental,mugler2016limits,bialek2005physical}. Eq.\ \ref{eq:ICvar2} assumes that the integration time is larger than the timescale of molecule diffusion over the radius $R$ of the collective,
$T \gg \tau_D = R^2/D$, though we relax this assumption in later simulations.
Following this procedure we find that $V$ and $C$ for IC are \cite{supinfo}
\begin{align}
    V_\text{IC} &= \frac{4\pi a^5c_0}{45DT} N  \label{eq:ICV} , \\
    C_\text{IC} &= -\frac{\pi a^5c_0}{18DT} \sum_{i \neq j}^N \frac{3\cos^2\Theta_{ij}-1}{n^3_{ij}} . \label{eq:ICC}
\end{align}
Here
$n_{ij}$
is the number of cell radii separating the centers of cells $i$ and $j$, and $\Theta_{ij}$ is the angle between the gradient direction and a line connecting the two cells.

$V_\text{IC}$ scales with $N$ since each cell is involved in gradient sensing.
However, Eq.\ \ref{eq:ICC} reveals an angular dependence on IC cell cross-correlations. A pair of cells can be correlated or anti-correlated depending on their locations relative to the gradient. Consider a pair of adjacent IC cells that are aligned parallel to the gradient ($\cos^2\Theta_{ij}=1$).
If a fluctuation causes an excess in chemoattractant near the boundary of the two cells, then the down-gradient cell will detect a molecule increase in its front half, resulting in an increased polarization; whereas the up-gradient cell will detect a molecule increase in its back half, resulting in a decreased polarization. The end result is an anti-correlation between the two cells. The opposite effect occurs if the two adjacent cells are aligned perpendicular to the gradient: fluctuations will affect both cells in the same way, causing positive correlations. Since contributions to $C_\text{IC}$ are dependent on cell pair locations, $C_\text{IC}$ itself will be dimensionality dependent because the angles made between cells are determined by geometry.

For a 1D chain of IC cells, every pair is parallel to the gradient resulting in anti-correlated measurements which we find in total scale as $N$ \cite{supinfo}. As dimensionality increases, more and more pairs of cells will be perpendicular to the gradient resulting in reduced anti-correlations in the collective. This culminates in 3D clusters having zero cell-cell covariance contribution to the total cluster variance \cite{supinfo}. This results in $\epsilon^2 \sim N^{-1}$ regardless of dimensionality, indicating that IC cells behave as effectively independent gradient sensors, even with diffusion-mediated cross-correlations. The scalings for $V$ and $C$ are summarized in Table \ref{table:1}. The resulting $\epsilon^2$ predictions are plotted in Fig.\ \ref{fig:2}(b) (blue lines), and we see excellent agreement with the simulations.

\begin{table}[t]
\normalsize
\begin{tabular}{cc|l|l|l|l|}
\cline{3-6}
& & $\langle P_z \rangle$ & $V$ & $C$ &
$\epsilon^2$ \\ \cline{1-6}
\multicolumn{1}{ |c| }{ }&&&&&\\[-1em]
\multicolumn{1}{ |c  }{\multirow{3}{*}{IC} } &
\multicolumn{1}{ |c| }{ 1D } & $N^{1}$ & $N^{1}$ & $-N^{1}$ & $N^{-1}$ \\
\multicolumn{1}{ |c| }{ }&&&&&\\[-1em]
\cline{2-6}
\multicolumn{1}{ |c| }{ }&&&&&\\[-1em]
\multicolumn{1}{ |c  }{} &
\multicolumn{1}{ |c| }{ 2D } & $N^{1}$ & $N^{1}$ & $-N^{1}$ & $N^{-1}$ \\
\cline{2-6}
\multicolumn{1}{ |c| }{ }&&&&&\\[-1em]
\multicolumn{1}{ |c  }{} &
\multicolumn{1}{ |c| }{ 3D } & $N^{1}$ & $N^{1}$ & $0$ & $N^{-1}$ \\
\hhline{|==|=|=|=|=|}
\multicolumn{1}{ |c| }{ }&&&&&\\[-1em]
\multicolumn{1}{ |c  }{\multirow{3}{*}{EC} } &
\multicolumn{1}{ |c| }{ 1D } & $N^1$ & $N^0$ & $-N^{-1}$ & $N^{-2}$ \\
\cline{2-6}
\multicolumn{1}{ |c| }{ }&&&&&\\[-1em]
\multicolumn{1}{ |c  }{} &
\multicolumn{1}{ |c| }{ 2D } & $N^1$ & $N^{1/2}$ & $N^{1/2}$ & $N^{-3/2}$\\
\cline{2-6}
\multicolumn{1}{ |c| }{ }&&&&&\\[-1em]
\multicolumn{1}{ |c  }{} &
\multicolumn{1}{ |c| }{ 3D } & $N^1$ & $N^{2/3}$ & $N^1$ & $N^{-1}$ \\
\cline{1-6}

\end{tabular}
\caption{Summary of scaling behavior. $N$ dependence of the leading order term for the mean $\langle P_z \rangle$, and the variance ($V$) and covariance ($C$) contributions to the relative error $\epsilon^2 = (V+C)/\langle P_z \rangle^2$. $C$ for EC in 2D has a $\log$ correction \cite{supinfo}.}
\label{table:1}
\end{table}

Next we turn our attention to EC, the mechanism in which grouped cells sense and migrate differently than individuals.
Often cells in a cluster differentiate, with edge cells polarized and bulk cells unpolarized \cite{malet2015collective,cai2016modeling}. In accordance with previous studies \cite{camley2016emergent,varennes2016collective},
we assume that cell interactions are mediated by contact inhibition of locomotion \cite{mayor2010keeping}. The interactions result in edge cells polarized away from their neighbors, and interior cells that remain uninvolved in chemical sensing and do not polarize [Fig.\ \ref{fig:1}(c)]. Since all the cells adhere to one another, the total polarization produced by the edge cells creates a motility force which acts on the whole collective. The edge cells polarize with strength proportional to the local concentration which, again like Berg and Purcell's perfect instrument \cite{berg1977physics}, is estimated by counting the molecules present within their cell volume. Hence we define the polarization of the $i$th cell in the collective as
\begin{equation} \label{eq:ECcell1}
    \vec{p}_i(t) =
    \begin{cases}
         \hat{r}_i \int_{U_i} d^3r \ c(\vec{r},t) &i \in \{ N_\text{edge} \} \\
        0 &i \in \{ N_\text{bulk} \} \ ,
    \end{cases}
\end{equation}
where $\hat{r}_i$ points radially outwards from the collective, and $U_i$ is the cell volume. Eq.\ \ref{eq:ECcell1} dictates that $\vec{p}_i$ is dependent on a cell's location relative to the collective. As illustrated in Fig.\ \ref{fig:1}(c), only the cells on the edge sense the chemoattractant, polarizing with a larger magnitude on the high concentration side of the collective, and the total polarization depends only on the cells along the edge:
$\vec{P} = \sum_{i \in \{N_\text{edge}\} } \vec{p}_i$.

Simulations for EC show that the mean polarization $\langle P_z \rangle$ scales with $N$ for all geometries [Fig.\ \ref{fig:2}(a), red points] even though $N_\text{edge}$ is dependent on the dimensionality of the collective. Our analytical solution helps us understand this result. For 1D EC, only the two opposing cells are polarized so $\langle\vec{P}\rangle$ can be solved for exactly, but for 2D and 3D we take the continuum limit of
$\vec{P} = \sum_i \vec{p}_i$,
assuming the collective is much larger than a single cell $R \gg a$ \cite{supinfo}. The resulting expressions are
\begin{equation} \label{eq:ECmean}
    \langle \vec{P} \rangle_\text{EC} = f_d a^4gN \ \hat{z},
\end{equation}
where the prefactors are $f_d = \{8\pi/3, 2\pi^2/3, 16\pi/9\}$ for $d=\{1,2,3\}$ dimensions, and for $d=1$ we have taken $N-1\to N$ for large $N$. Eq.\ \ref{eq:ECmean} is shown in Fig.\ \ref{fig:2}(b) (red lines), and we see good agreement. $\langle P_z \rangle$ scales with $N$ because it depends on the product of
$N_\text{edge} \sim N^{(d-1)/d}$
and the distance spanned in the gradient direction
$R \sim N^{1/d}$,
resulting in a mean polarization which is geometry invariant \cite{malet2015collective}.

Comparing EC and IC shows that $\langle P_z \rangle \sim g$ in both cases, which is consistent with some single-cell chemotaxis experiments \cite{kim2013cooperative,wang2004differential}, and that
$\langle P_z \rangle \sim N$ regardless of collective migration mechanism or geometry as seen in Fig.\ \ref{fig:2}(a).
The only difference is that
$\langle P_z \rangle_\text{EC} \approx 6 \langle P_z \rangle_\text{IC}$,
although this relatively small difference may be difficult to detect in biological systems.
Does the same equivalence between EC and IC also hold for the relative error?

Interestingly, simulations show that the EC relative error does depend on geometry and in fact outperforms IC in terms of scaling in 1D and 2D [Fig.\ \ref{fig:2}(b), red points]. Only in 3D does the relative error appear to scale the same as IC. In order to understand the dimension dependence of the EC relative error we again investigate the model analytically.
Following the procedure outlined by Eqs.\ \ref{eq:ICvar1} and \ref{eq:ICvar2} we find analytic expressions for $\text{Var}[P_z] = V + C$ for EC \cite{supinfo},
\begin{align}
    V_\text{EC} &= \frac{16\pi a^5c_0}{15DT} \sum_{i=1}^{N_\text{edge}} \cos^2\Theta_i \label{eq:ECV} , \\
    C_\text{EC} &= \frac{8\pi a^5c_0}{9DT} \sum_{i \neq j}^{N_\text{edge}} \frac{\cos\Theta_i\cos\Theta_j}{n_{ij}} , \label{eq:ECC}
\end{align}
with $\Theta_i$ the angle $\hat{r}_i$ makes with the gradient. Both $V_\text{EC}$ and $C_\text{EC}$ depend on dimensionality simply because $N_\text{edge} \sim N^{(d-1)/d}$. From Eqs.\ \ref{eq:ECV} and \ref{eq:ECC} we see that
$V \sim N_\text{edge}$,
and that $C$ depends on the angles edge cells make with the gradient. The angular dependence means that cells along the front and back sides of the cluster (relative to the gradient) are anti-correlated since $\cos\Theta_i\cos\Theta_j \approx -1$,
whereas pairs of edge cells near the middle are very weakly correlated ($\cos\Theta_i\cos\Theta_j \approx 0$). Unlike in the case of IC, the scaling of $C$ with $N$ increases with dimensionality \cite{supinfo} as summarized in Table \ref{table:1}, and the resulting $\epsilon^2$ predictions show good agreement with the simulation results [Fig.\ \ref{fig:2}(b)].

The dimension dependence of the EC relative error can be understood by thinking of the collective as one large detector whose sensory surface is comprised of two halves. If both halves were to take measurements of their local concentrations and then polarize in opposing directions with strengths proportional to their measurements, then $\epsilon^2$ would depend inversely on the radius of each half
$a_\text{eff}$ \cite{berg1977physics}
and inversely on the square of their separation distance
$A_\text{eff}$: $\epsilon^2  \sim a^{-1}_\text{eff}A^{-2}_\text{eff}$ \cite{mugler2016limits}.
The radius of each half is independent of $N$ for a 1D chain (each half is a single cell), but it scales as
$a_\text{eff} \sim N^{1/d}$ for $d = 2$ or $3$ dimensions.
The separation distance scales with the radius of the collective for all $d$,
$A_\text{eff} \sim N^{1/d}$. This results in
$\epsilon^2 \sim \{N^{-2}, N^{-3/2}, N^{-1}\}$ for $d=\{1,2,3\}$ [Fig.\ \ref{fig:2}(b), black lines],
which agree with the scalings seen in simulations and analytics.

Thus, the physical origin of the advantage of EC over IC lies in how the errors scale with the collective size $N$. In IC, all $N$ cells contribute to the sensing, and cross-correlations between them scale either linearly or sublinearly with $N$, leading to a scaling $\epsilon \sim 1/\sqrt{N}$ that is characteristic of independent sensors. But in EC, only $N_\text{edge} \sim N^{(d-1)/d}$ cells contribute to the sensing, leading to a sublinear scaling with $N$ of the variance contributions of the individual cells. The total variance of the collective, then, depends on the cross-correlations, which are geometry-specific: in 1D they are dwarfed by the individual variances, in 2D they are commensurate, and in 3D they dominate (Table \ref{table:1}). As a result, 1D and 2D EC collectives benefit from a variance that scales subextensively, i.e., sublinearly with $N$.

Our analytical treatment relies on several assumptions which we now relax using the simulations. In Fig.\ \ref{fig:2} the integration time $T$ is larger than the timescale for molecule diffusion $\tau_D$.
We find that reducing $T$ to as much as two orders of magnitude smaller than $\tau_D$ still results in the expected relative error scalings, with 1D and 2D EC outperforming IC \cite{supinfo}. Similarly, using an exponential concentration profile instead of a linear one does not change the relative error scaling behavior \cite{supinfo}.

In our model, IC polarization is adaptive to the background concentration as observed in the Ras signaling pathway for \textit{Dictyostelium discoideum} chemotaxis \cite{takeda2012incoherent}. On the other hand, our EC model is non-adaptive. Cell polarization increases with background concentration causing tension in the collective [Fig.\ \ref{fig:1}(c)], as previously studied \cite{camley2016emergent}. However, adaptive collective sensing has been observed in mammary epithelial cells \cite{ellison2016cell}. Our EC model could be made adaptive by replacing the integrand in Eq.\ \ref{eq:ECcell1} with
$c(\vec{r}+\vec{r}',t)-c_0$
assuming that information about the chemical can be reliably conveyed across the collective.
This change does not affect the properties of $\vec{P}$ since the background concentration cancels out when summing over all edge cells, but it does remove the internal tension in the collective.

Besides the advantage revealed here in terms of chemotactic precision, there may be other natural advantages to EC. In EC only edge cells are involved in chemical sensing and polarization, freeing bulk cells from receptor and protein production necessary for chemotaxis. Bulk cells are free to differentiate into other phenotypes, which is in stark contrast with IC where every cell must be of the polarized phenotype. Additionally, EC provides a simple solution to bulk cells being shielded from the diffusing chemoattractant by edge cells. This phenomenon is especially important for 3D collectives where it can significantly impact the sensory precision of bulk cells \cite{smith2016role}.

The above advantages may be why EC-style collective migration is more prevalent than IC. For example, EC has been observed in two dimensional collectives of malignant lymphocytes \cite{malet2015collective} and in border cell migration \cite{cai2016modeling}. In cancer, metastatic invasion sometimes occurs in the form of chains of cells leaving the tumor with a leader cell at the front \cite{cheung2013collective,friedl2009collective}, analogous to our 1D EC model.
Two-dimensional EC migration may also be implicated in tumorigenesis and metastasis in pancreatic ductal cells \cite{bardeesy2002pancreatic} given the cylindrical surface-like geometry of pancreas ducts
\cite{footnote4}.

How can our predictions be tested in experiments? The chemotactic index (CI), commonly defined as
$\text{CI} \equiv \langle \cos\theta \rangle$ where $\theta$ is the angle between the trajectory and the gradient \cite{van2007biased}, is actually a simple monotonic function of $\epsilon^2$. For small deviations from perfect chemotaxis, we have
$\text{CI} \approx 1 - \langle \theta^2 \rangle/2 = 1 - \text{Var}[\theta]/2$.
If $v_z$ and $v_x$ are the components of the velocity of the collective parallel and perpendicular to the gradient, respectively, then $\theta \approx v_x/v_z$ with $\langle v_z\rangle > 0$ and $\langle v_x\rangle = 0$, resulting in
$\text{Var}[\theta] = \text{Var}[v_z] / \langle v_z\rangle^2 = \epsilon^2$.
Therefore the relative error and chemotactic index are related as
$\text{CI} = 1 - \epsilon^2/2$ for small errors.
With this relationship the predicted scalings of $\epsilon^2$ for EC and IC may be tested experimentally. Estimates for the relative error of two example biological systems are provided in the supplement \cite{supinfo}.

We have shown how the fluctuations in a diffusing attractant concentration set physical limits to collective chemotactic performance. By focusing on two fundamental classes of collective chemotaxis, we have found that the mean polarization scales with the size of the collective irrespective of the mechanism or geometry, but that the emergent mechanism outperforms an individual-based one for 1D and 2D geometries in terms of chemotactic precision.
This advantage arises due to the ways that errors accumulate in the two mechanisms: in an emergent strategy, fewer cells contribute their sensory noise to the collective, and in 1D and 2D the cross-correlations between cells remain low, ultimately leading to a subextensive scaling of polarization variance with collective size. As such, the performance advantage is an inherent property of the emergent mechanism, and we suspect that it not only helps explain the prevalence of emergent chemotaxis in cellular systems, but that it also is detectable using standard measures such as the chemotactic index.

This work was supported by the Ralph W.\ and Grace M.\ Showalter Research Trust, the Purdue Research Foundation, the Purdue University Center for Cancer Research Challenge Award, and Simons Foundation Grant No.\ 376198.


\newpage

\widetext
\begin{center}
\textbf{\large Supplementary Information for \\ ``Emergent versus Individual-based Multicellular Chemotaxis''}
\end{center}

\renewcommand{\thefigure}{S\arabic{figure}}
\renewcommand{\theequation}{S\arabic{equation}}
\setcounter{equation}{0}
\setcounter{figure}{0}

\onecolumngrid

\section{Generality of the IC and EC Models}
In this section we discuss how various methods of collective chemotaxis studied in the literature compare to the IC and EC models presented in the main text.

Known strategies of collective cell chemotaxis fall broadly into five categories. First, experiments focused on contact inhibition of locomotion have revealed collective cell streaming in which each cell makes independent protrusions (Ref.\ [32] of the main text). Second, Refs.\ [2,7,12,13] of the main text all show behavior wherein edge cells exhibit an active, motile phenotype or make outward protrusions. Third, Refs.\ [1,8,19] of the main text show behavior wherein active, motile cells form one or more multicellular chain-like protrusions extending from the collective. Fourth, experiments on epithelial organoids have demonstrated that chemical communication between cells can underlie collective gradient sensing (Ref.\ [5] of the main text). Finally, recent modeling studies have highlighted the role played by cell rearrangement within the collective in governing collective chemotaxis (Refs.\ [16,39] of the main text).

The first and second strategies are directly described by our IC and EC models, respectively. The third strategy may be considered as a combination of our IC and EC models: the cell at the tip of the multicellular protrusion is often of a highly invasive phenotype akin to our EC edge cells, while the cells within the bulk of the protrusion are less invasive and may behave like bulk cells or IC cells depending on their activity. In the case of the fourth strategy, the error in the communication process will contribute additional noise to the collective polarization (Ref. [31] of the main text), and when communication is optimal we recover the same scaling relationship for the relative error as in our EC model, as discussed in the second paragraph on page 4 of the main text. Finally, the fifth strategy, namely collective chemotaxis in which cells rearrange, is not encompassed by our IC and EC models, since we consider cell locations to be fixed relative to the collective. It may very well be that cell rearrangement allows for spatial fluctuations, and thereby correlations, to be averaged out resulting in a quantitatively improved relative error. This is an interesting avenue of further research.

\section{Particle-based Simulations}

We perform computational simulations in order to test the properties of EC and IC for one, two and three dimensional collectives. In the simulation, particles move randomly inside a 3D volume and boundaries are set to produce the desired concentration profiles. Cells are placed at fixed positions in the 3D volume in either one, two or three dimensional configurations. For a linear concentration profile, one boundary produces particles, and the opposing one is an absorbing boundary while all other boundaries periodic. For an exponential concentration, the same boundaries are used and particles are also allowed to degrade.

At each time-step of the simulation particles randomly move and are produced. In a given time-step particles move in a random direction with a probability
$p = D\Delta t/b^2$,
with $b$ the particle hopping distance, $D$ the diffusion coefficient, and $\Delta t$ the time-step. A particle is produced during that time-step with probability
$q = k\Delta t$,
with $k$ the production rate. The time-step $\Delta t$ is set such that $p + q \leq 1$. In the case of an exponential concentration profile, particles may also degrade during a time-step. Particles degrade with probability
$r = \beta \Delta t$, with $\beta$ the degradation rate. In this case $\Delta t$ is set such that $p + q + r \leq 1$.

The simulation code used for this paper can be found at \url{https://doi.org/10.5281/zenodo.401040}, and the most up-to-date version of the code can be found at \url{https://github.com/varennes/particletrack}.

\section{Analytic Results}

We consider collectives in one, two and three dimensions of radius $R$ comprised of $N$ cells. Each cell is taken to be a permeable sphere of radius $a$ through which molecules of the surrounding chemical concentration $c(\vec{r},t)$ can freely diffuse. The chemical concentration is taken to be
\begin{equation}
    c(\vec{r},t) = c(0,t) + \vec{r}\cdot\vec{g}(\vec{r},t)
\end{equation}
with $\vec{g}$ parallel to the \textit{z} axis. The chemical concentration obeys normal diffusion
\begin{equation} \label{eq:diffeq}
    \dot{c} = D\nabla^2c+\eta_c
\end{equation}
with $D$ the diffusion coefficient, and $\eta_c$ the Langevin noise due to fluctuations in concentration. We linearize the concentration $c(\vec{r},t) = \bar{c}(\vec{r}) + \delta c(\vec{r},t)$ with
\begin{equation} \label{eq:meanc}
    \bar{c}(\vec{r}) = c_0 + \vec{r}\cdot\vec{g}
\end{equation}
where $c_0$ is the mean concentration at the origin. The Langevin noise term $\eta_c$, and the Fourier transformed fluctuation in the concentration $\delta\tilde{c}(\vec{k},\omega)$ have the following properties (see Ref.\ [23] of the main text):
\begin{gather}
    \langle\tilde{\eta}_c(\vec{k}',\omega')\tilde{\eta}_c(\vec{k},\omega) \rangle = 2D \ 2\pi\delta(\omega-\omega') \int d^3y \ \vec{k}\cdot\vec{k}' \ \bar{c}(\vec{y}) \ e^{i\vec{y}\cdot\left(\vec{k}-\vec{k}'\right)} \ ,
    \label{eq:c1} \\
    \langle\delta\tilde{c}(\vec{k}',\omega')\delta\tilde{c}(\vec{k},\omega) \rangle = \frac{\langle\tilde{\eta}_c(\vec{k}',\omega')\tilde{\eta}_c(\vec{k},\omega) \rangle}{(Dk^2-i\omega)(Dk'^2+i\omega')} \ .
    \label{eq:c2}
\end{gather}

Next, we define the cell polarization vectors for individual-based chemotaxis (IC) and emergent chemotaxis (EC). Collectives of $N$ cells form shapes of different dimensionality: a chain of cells of length $2R$ (1D), a disc of cells with radius $R$ (2D), and a sphere of cells of radius $R$ (3D).

\subsection{Individual-based Chemotaxis}

In the IC mechanism, cells independently measure the chemoattractant gradient in order to set their polarization vector $\vec{p}$. For a spherically-shaped cell with volume $U_i$, $\vec{p}_i$ is defined as
\begin{equation}
    p_{i\alpha}(t) = \int_{U_i} d^3r \ w_\alpha c(\vec{r},t) ,
\end{equation}
where $\alpha\in\{x,y,z\}$, and in spherical coordinates the cosine is $w_\alpha = \{\sin\theta \cos\phi, \sin\theta \sin\phi, \cos\theta\}$.
The $x, y, z$ components are written as
\begin{align}
    p_{ix}(t) &= \int d\Omega' \ \sin\theta' \cos\phi' \int_0^a dr' \ r'^2 \ c(\vec{r}_i+\vec{r}',t)  \label{eq:3DMWpx} \\
    p_{iy}(t) &= \int d\Omega' \ \sin\theta' \sin\phi' \int_0^a dr' \ r'^2 \ c(\vec{r}_i+\vec{r}',t)  \label{eq:3DMWpy} \\
    p_{iz}(t) &= \int d\Omega' \ \cos\theta' \int_0^a dr' \ r'^2 \ c(\vec{r}_i+\vec{r}',t), \label{eq:3DMWpz}
\end{align}
where $d\Omega' = \sin\theta'd\theta'd\phi'$. The $r'$ coordinates are relative to the center of the respective cell, and the $r_i$ coordinates are relative to the center of the collective.
Using the mean concentration (Eq.\ \ref{eq:meanc}) with a constant gradient $\vec{g} = g \hat{z}$, we calculate the mean polarization of a single cell:
\begin{align*}
    \langle p_{iz} \rangle &= \int d\Omega' \ \cos\theta' \int_0^a dr' \ r'^2 \ \bar{c}(\vec{r}_i+\vec{r}') \\
    &= \int_0^a dr' \ r'^2 \int d\Omega' \ \cos\theta' (c_0+gr_i\cos\theta_i+gr'\cos\theta') \\
    &= \frac{\pi}{3} a^4 g \ .
\end{align*}
The means for the $x$ and $y$ components are $\langle p_{ix} \rangle = \langle p_{iy} \rangle = 0$ since they are perpendicular to the gradient. On average, cells performing IC migration will only polarize in the $z$ direction. The mean for a collective of IC cells is
\begin{equation}
    \langle \vec{P} \rangle = \frac{\pi}{3} a^4 g N \ \hat{z} \ .
\end{equation}

\subsection{Emergent Chemotaxis}

In EC, cells along the edge of the cluster polarize outwards, whereas cells in the interior are not involved in chemical sensing and remain unpolarized:
\begin{equation}
    \vec{p}_i(t) =
    \begin{cases}
         \hat{r}_i \int_{U_i} d^3r \ c(\vec{r},t) &i \in \{ N_\text{edge} \} \\
        0 &i \in \{ N_\text{bulk} \} \ ,
    \end{cases}
\end{equation}
where $\hat{r}$ points radially outwards from the collective. In order to break down $\vec{p}_i(\vec{r},t)$ into component vectors we must be mindful of the dependence of $\hat{r}_i$ on the cell location. For an edge cell the unit vector $\hat{r}_i$ points in the direction of the cell's location in the collective,
$\hat{r}_i = \sin\Theta_i\cos\Phi_i \hat{x} + \sin\Theta_i\sin\Phi_i \hat{y} + \cos\Theta_i \hat{z}$
where $\Theta_i$ is the polar angle made with the gradient direction and $\Phi_i$ is the azimuthal angle along the collective. The cell component vectors are
\begin{align}
    p_{ix}(t) &= \sin\Theta_i\cos\Phi_i \ p_i(t) , \\
    p_{iy}(t) &= \sin\Theta_i\sin\Phi_i \ p_i(t) \ , \\
    p_{iz}(t) &= \cos\Theta_i \ p_i(t) \ ,
\end{align}
with $p_i(t) = \int_{U_i} d^3r \ c(\vec{r},t)$ and $i \in \{ N_\text{edge}\}$. The total polarization of the collective,
$\vec{P} = \vec{P}_x + \vec{P}_y + \vec{P}_z$,
is a sum of all the component vectors:
\begin{align}
    P_x(t) &= \sum_i^{N_\text{edge}} \sin\Theta_i\cos\Phi_i \ p_i(t) \ , \label{eq:PECx} \\
    P_y(t) &= \sum_i^{N_\text{edge}} \sin\Theta_i\cos\Phi_i \ p_i(t) \ , \label{eq:PECy} \\
    P_z(t) &= \sum_i^{N_\text{edge}} \cos\Theta_i \ p_i(t) \ . \label{eq:PECz}
\end{align}

For an edge cell, the mean polarization is equal to the average number of molecules the cell counts within its spherical body:
\begin{align*}
    \langle \vec{p}_i \rangle &= \int_{U_i} d^3r \ \bar{c}(\vec{r}) \ \hat{r}_i \\
    &= \frac{4\pi}{3}a^3 (c_0+gR\cos\Theta_i) \ \hat{r}_i \ ,
\end{align*}
where $\Theta_i$ is the angle the cell's location makes with the gradient direction. The mean for a cluster of EC cells will depend on the dimensionality of the cluster. For a 1D chain of cells, only the two cells on the opposite ends of the chain are polarized, and $\langle P \rangle$ is the difference in the mean number of molecules counted in between the two edge cells:
\begin{equation} \label{eq:1DECmean}
    \text{1D}: \langle \vec{P} \rangle = \frac{8\pi}{3}a^4g (N-1) \ \hat{z} \ .
\end{equation}

In order to calculate the mean total polarization for two and three dimensional clusters we assume that the cluster size is relatively large ($a\ll R$) and approximate the sum as an integral. For a 2D disc of cells the sum
$\vec{P} = \sum_i^{N_\text{edge}} \vec{p}_i$
becomes an integral over the circumference of the cluster. The circumference and the total number of cells along the edge are related by $2\pi R = 2a N_\text{edge}$, and so a segment along the perimeter of length $R\theta$ is equivalent in length to $2a n$ with $n$ the number of edge cells in that segment. Hence $n =\frac{R}{2a}\theta$ allowing us to write integrals for $\vec{P}(t)$ as
\begin{equation}
    \vec{P}(t) = \frac{R}{2a} \int_0^{2\pi} d\theta \ \vec{p}_i(t) \ .
\end{equation}
The mean polarization will point only in the $z$ direction with magnitude
\begin{equation*}
    \langle P_z \rangle = \frac{R}{2a} \int_0^{2\pi} d\theta \ \langle p_z \rangle = \frac{R}{2a} \int_0^{2\pi} d\theta \ \cos\theta \left( \frac{4\pi}{3}a^3 (c_0+gR\cos\theta) \right) = \frac{2\pi^2}{3} a^2gR^2 \ .
\end{equation*}
Using the relation $N = (R/a)^2$, the mean of the total polarization is
\begin{equation} \label{eq:2DECmean}
    \text{2D}: \langle\vec{P}\rangle = \frac{2\pi^2}{3} a^4gN \ \hat{z} \ .
\end{equation}

Similarly, in 3D we approximate the sum as an integral of the spherical surface of the cluster. A patch on the surface of area $\Omega R^2$ encompasses
$n = \Omega R^2 / (\pi a^2)$ edge cells. The total polarization can therefore be written as an integral over the surface of a spherical cluster:
\begin{equation}
    \vec{P}(t) = \frac{R^2}{\pi a^2} \int d\Omega \ \vec{p}_i(t) \ .
\end{equation}
The mean polarization will point only in the $z$ direction with magnitude
\begin{equation*}
    \langle P_z \rangle = \frac{R^2}{\pi a^2} \int d\Omega \ \langle p_z \rangle = \frac{R^2}{\pi a^2} \int d\Omega \ \cos\theta \left( \frac{4\pi}{3}a^3 (c_0+gR\cos\theta) \right) = \frac{16\pi}{9} agR^3 \ .
\end{equation*}
For a spherical cluster, $N = (R/a)^3$ and the mean of the total cluster polarization is
\begin{equation} \label{eq:3DECmean}
    \text{3D}: \langle\vec{P}\rangle = \frac{16\pi}{9} a^4gN \ \hat{z} \ .
\end{equation}

\subsection{Variance in Cell \& Cluster Polarization}

Here we derive the variance in cell and collective polarizations. The first section gives a general outline for how this is done for either collective migration mechanism. The following sections will derive the specific expressions for IC, and EC and provide scaling arguments for 1D, 2D and 3D geometries.

\subsubsection{General Outline}
Since the total collective polarization is a sum of the cell polarization for IC or EC, the variance in the total polarization takes the general form:
\begin{equation} \label{eq:PVAR1}
\begin{split}
    \text{Var}[P_\alpha] &=
    \underbrace{\sum_{i=1}^N \text{Var}[p_{i,\alpha}]}_\text{variance contribution}
    + \underbrace{\sum_{i\neq j} \text{Cov}[p_{i,\alpha},p_{j,\alpha}]}_\text{covariance contribution} \\
    &\equiv V + C \ ,
\end{split}
\end{equation}
with $\alpha \in \{ x,y,z\}$.
In order to derive an expression for the variance in collective polarization we must first understand the fluctuations occurring in single cell measurements. The fluctuations in the $i^\text{th}$ cell's polarization vector are calculated by linearizing each component,
$p_{i,\alpha}(t) = \bar{p}_{i,\alpha} + \delta p_{i,\alpha}(t)$
and taking the Fourier transform. The Fourier transform of $\delta p_{i,\alpha}(t)$ takes the general form
\begin{equation}
    \delta\tilde{p}_{i,\alpha}(\omega) = \int d^3x_i \int \frac{d^3k}{(2\pi)^3} f(\theta_i,\phi_i) \ \delta\tilde{c}(\vec{k},\omega) \ e^{-i\vec{k}\cdot(\vec{x}_i+\vec{x})}
\end{equation}
where the functional form of $f(\theta_i,\phi_i)$ is dictated by the migration mechanism (EC or IC) and the component $\alpha$.
The cross-spectrum between the $i^\text{th}$ and $j^\text{th}$ cells is
$\langle \delta\tilde{p}_{i,\alpha}^*(\omega') \delta\tilde{p}_{j,\alpha}(\omega) \rangle$. Utilizing the cross-spectrum we can derive an expression for the variance and covariance in the long-time averaged cell polarization by calculating the power spectrum
\begin{equation} \label{eq:sij1}
    S_{ij,\alpha}(\omega=0) = \lim_{\omega \to 0} \int \frac{d\omega'}{2\pi}
    \langle \delta\tilde{p}_{i,\alpha}^*(\omega') \delta\tilde{p}_{j,\alpha}(\omega) \rangle \ .
\end{equation}
The cell polarization variance and covariance is given by:
\begin{align}
    \text{Var}[p_{i,\alpha}] &= \frac{1}{T} S_{ii,\alpha}(0) \ , \\
    \text{Cov}[p_{i,\alpha},p_{j,\alpha}] &= \frac{1}{T} S_{ij,\alpha}(0) \ ,
\end{align}
where $T$ is the averaging time.
With the above expressions for the cell polarization variance and covariance we can solve for Eq.\ \ref{eq:PVAR1} and in turn calculate the relative error for the whole collective. In subsequent sections we show the derivation only for the $z$ component of the polarization since it is parallel to the gradient. The expressions $x$ and $y$ components will be equal to to the $z$ component since the fluctuations in concentration are symmetric in each direction.

\subsection{Individual-based Chemotaxis}
For IC the variance in $P_z$ is
\begin{equation}
    \text{Var}[P_z] = \sum_{i=1}^N \text{Var}[p_{i,z}] + \sum_{i\neq j} \text{Cov}[p_{i,z},p_{j,z}] \equiv V_\text{IC} + C_\text{IC} \ .
\end{equation}
The Fourier-transformed fluctuations in IC cell polarization is
\begin{equation}
    \delta\tilde{p}_{j,z}(\vec{k},\omega) = \int_V d^3x \int \frac{d^3k}{(2\pi)^3} \cos\theta
    \delta\tilde{c}(\vec{k},\omega)
    e^{-i\vec{k}\cdot(\vec{x_i}+\vec{x})} \ .
\end{equation}
The cross-spectrum for the $z$-component between two cells is
\begin{equation} \label{eq:pij1}
\begin{split}
    \langle\delta\tilde{p}_{i,z}^*(\omega')\delta\tilde{p}_{j,z}(\omega) \rangle = \int_V d^3x d^3x' \int \frac{d^3kd^3k'}{(2\pi)^6}
    \cos\theta \cos\theta'
    \langle \delta\tilde{c}^*(\vec{k}',\omega')\delta\tilde{c}(\vec{k},\omega) \rangle
    e^{-i\vec{k}\cdot(\vec{x}_j+\vec{x})} e^{i\vec{k}'\cdot(\vec{x}_i+\vec{x}')} \ .
\end{split}
\end{equation}
We can rewrite Eq.\ \ref{eq:pij1} by noting that only the relative locations of cell $i$ and $j$ are relevant for the cross-spectrum. Let $\vec{r}_{ij} = \vec{x}_i - \vec{x}_j$ and $r_{ij} = |\vec{r}_{ij}|$.
\begin{equation} \label{eq:sij2}
\begin{split}
    \langle \delta\tilde{p}_{i,z}^*(\omega')\delta\tilde{p}_{j,z}(\omega) \rangle = \int_V d^3x d^3x' \int \frac{d^3kd^3k'}{(2\pi)^6}
    \cos\theta \cos\theta' \\
    \langle \delta\tilde{c}^*(\vec{k}',\omega') \delta\tilde{c}(\vec{k},\omega) \rangle
    e^{-i\vec{k}\cdot\vec{x}} e^{i\vec{k}'\cdot(\vec{r}_{ij}+\vec{x}')}
\end{split}
\end{equation}
Plugging in Eq.\ \ref{eq:c2} for
$\langle \delta\tilde{c}^*(\vec{k}',\omega') \delta\tilde{c}(\vec{k},\omega) \rangle$
and writing $\cos\theta$ in terms of spherical harmonic $Y_1^0(\hat{x})$ yields
\begin{align*}
    \langle\delta\tilde{p}_{i,z}^*(\omega')\delta\tilde{p}_{j,z}(\omega) \rangle &= \int_V d^3x d^3x' \int \frac{d^3kd^3k'}{(2\pi)^6}
    \frac{4\pi}{3} Y_1^0(\hat{x})Y_1^0(\hat{x}') \ 2D \\ &\frac{2\pi\delta(\omega-\omega')}{(Dk^2-i\omega)(Dk'^2+i\omega')} \int d^3y \vec{k}\cdot\vec{k}' \bar{c}(\vec{y}) e^{i\vec{y}\cdot(\vec{k}-\vec{k}')}
    e^{-i\vec{k}\cdot\vec{x}} e^{i\vec{k}'\cdot(\vec{r}_{ij}+\vec{x}')} \\
    &= \frac{4D}{3(2\pi)^5} 2\pi\delta(\omega-\omega') \int_V d^3x d^3x' \int d^3kd^3k'd^3y \ Y_1^0(\hat{x})Y_1^0(\hat{x}') \\
    &\frac{\bar{c}(\vec{y}) \ \vec{k}\cdot\vec{k}'\ e^{i\vec{y}\cdot(\vec{k}-\vec{k}')}}{(Dk^2-i\omega)(Dk'^2+i\omega')} e^{-i\vec{k}\cdot\vec{x}} e^{i\vec{k}'\cdot(\vec{r}_{ij}+\vec{x}')}
\end{align*}
Plugging in the above expression for
$\langle\delta\tilde{p}_{i,z}^*(\omega')\delta\tilde{p}_{j,z}(\omega) \rangle$
into $S_{ij,z}(0)$ (Eq.\ \ref{eq:sij1}) and using the specified mean concentration from Eq.\ \ref{eq:meanc}:
\begin{equation}
\begin{split}
    S_{ij,z}(0) = \frac{4}{3(2\pi)^5D} \int_V d^3x d^3x' \int d^3kd^3k'd^3y \ Y_1^0(\hat{x})Y_1^0(\hat{x}') \frac{\vec{k}\cdot\vec{k}'}{k^2k'^2} \\
    (c_0+\vec{g}\cdot\vec{y}) \ e^{i\vec{y}(\vec{k}-\vec{k}')} \ e^{-i\vec{k}\cdot\vec{x}} e^{i\vec{k}'\cdot(\vec{r}_{ij}+\vec{x}')} \ .
\end{split}
\end{equation}
We can break up the expression for $S_{ij,z}(0)$ into two terms: one dependent on the background concentration, the other on the gradient.
\begin{equation} \label{eq:sij3}
\begin{split}
    S_{ij,z}(0) = \frac{4}{3(2\pi)^5D} \int_V d^3x d^3x' \int d^3kd^3k' \ Y_1^0(\hat{x})Y_1^0(\hat{x}') \frac{\vec{k}\cdot\vec{k}'}{k^2k'^2} \ e^{-i\vec{k}\cdot\vec{x}} \\
    e^{i\vec{k}'\cdot(\vec{r}_{ij}+\vec{x}')}
    \left((2\pi)^3 \delta^3(\vec{k}-\vec{k}') c_0 + \int d^3y \ \vec{g}\cdot\vec{y} \ e^{i\vec{y}(\vec{k}-\vec{k}')} \right)
\end{split}
\end{equation}
Let $S_{ij}^1$ represent the background concentration term and $S_{ij}^2$ represent the gradient dependent term in the power spectrum such that $S_{ij,z}(0) = S_{ij}^1 + S_{ij}^2$.
\begin{equation} \label{eq:Sij1MW}
    S_{ij}^1 = \frac{4c_0}{3(2\pi)^2D} \int_V d^3x d^3x' \int d^3k \ Y_1^0(\hat{x})Y_1^0(\hat{x}') \frac{1}{k^2} \ e^{-i\vec{k}\cdot\vec{x}} \
    e^{i\vec{k}\cdot(\vec{r}_{ij}+\vec{x}')}
\end{equation}
\begin{equation} \label{eq:Sij2MW}
\begin{split}
    S_{ij}^2 = \frac{4c_0}{3(2\pi)^5D} \int_V d^3x d^3x' \int d^3kd^3k'd^3y \ Y_1^0(\hat{x})Y_1^0(\hat{x}') \frac{\vec{k}\cdot\vec{k}'}{k^2k'^2} \ \vec{g}\cdot\vec{y} \\
    e^{i\vec{y}(\vec{k}-\vec{k}')} \
    e^{-i\vec{k}\cdot\vec{x}} \
    e^{i\vec{k}'\cdot(\vec{r}_{ij}+\vec{x}')}
\end{split}
\end{equation}
The following expansions will prove useful:
\begin{align}
    e^{-i\vec{k}\cdot\vec{r}} &= 4\pi \sum_{l,m} (-i)^l j_l(kr) Y_l^{m}(\hat{k}) Y_l^{m*}(\hat{r}) \ , \label{eq:planewave} \\
    \vec{a}\cdot\vec{b} &= \frac{4\pi}{3} ab \sum_{m=-1}^1 Y_1^m(\hat{a}) Y_1^{m*}(\hat{b}) \ . \label{eq:dotprod}
\end{align}
Starting with Eq.\ \ref{eq:Sij1MW} we expand all the exponential terms, and we use these expansions in order to evaluate the angular integrals in $S_{ij}^1$.
\begin{equation}
\begin{split}
    S_{ij}^1 = \frac{2^5(2\pi)c_0}{3D} \int_V d^3x d^3x' \int d^3k \ Y_1^{0*}(\hat{x})Y_1^{0*}(\hat{x}') \ \frac{1}{k^2} \\
    \left(\sum_{l_1,m_1}i^{-l_1} j_{l_1}(xk) Y_{l_1}^{m_1}(\hat{x}) Y_{l_1}^{m_1*}(\hat{k}) \right)
    \left(\sum_{l_2,m_2}i^{l_2} j_{l_2}(r_{ij}k) Y_{l_2}^{m_2}(\hat{k}) Y_{l_2}^{m_2*}(\hat{r}_{ij}) \right) \\
    \left(\sum_{l_3,m_3}i^{l_3} j_{l_3}(x'k) Y_{l_3}^{m_3}(\hat{k}) Y_{l_3}^{m_3*}(\hat{x}') \right)
\end{split}
\end{equation}
The angular integrals over $\hat{x}$ and $\hat{x}'$ eliminate the summations over $l_1,m_1$ and $l_3,m_3$.
\begin{equation} \label{eq:Sij3MW}
\begin{split}
    S_{ij}^1 = \frac{2^5(2\pi)c_0}{3D} \int_0^a dx dx' \int d^3k \ \frac{1}{k^2} x^2 x'^2 \ j_{1}(xk) j_{1}(x'k) \ Y_{1}^{0*}(\hat{k}) Y_{1}^{0}(\hat{k}) \\
    \left(\sum_{l_2,m_2}i^{l_2} j_{l_2}(r_{ij}k') Y_{l_2}^{m_2}(\hat{k}') Y_{l_2}^{m_2*}(\hat{r}_{ij}) \right)
\end{split}
\end{equation}
The product of the two spherical harmonics is
\begin{equation*}
    Y_{1}^{0*}(\hat{k}) Y_{1}^{0*}(\hat{k}) = \frac{1}{\sqrt{4\pi}} \left( Y_0^0(\hat{k}) + \frac{2\sqrt{5}}{5} Y_2^0(\hat{k}) \right) \ .
\end{equation*}
Therefore when evaluating the $\hat{k}$ integral in Eq.\ \ref{eq:Sij3MW} only the $l_2=0,m_2=0$ and $l_2=2,m_2=0$ terms of the summation will be non-zero.
\begin{equation}
\begin{split}
    S_{ij}^1 = \frac{2^5(2\pi)c_0}{3D\sqrt{4\pi}} \int_0^a dx dx' \int_0^\infty dk \ x^2 x'^2 \ j_{1}(xk) j_{1}(x'k) \\ \left(j_0(r_{ij}k) Y_0^0(\hat{r}_{ij}) - \frac{2\sqrt{5}}{5} j_2(r_{ij}k) Y_2^0(\hat{r}_{ij}) \right)
\end{split}
\end{equation}
The integrals over $x$ and $x'$ evaluate to:
\begin{equation*}
    \int_0^a dx \ x^2 j_1(kx) = \frac{1}{k^3}\left(2-2\cos(ak)-ak\sin(ak) \right) \equiv \frac{1}{k^3} h(ak) \ .
\end{equation*}
Note that
$Y_0^0(\Theta_{ij},\Phi_{ij}) = \frac{1}{\sqrt{4\pi}}$
, and
$Y_2^0(\Theta_{ij},\Phi_{ij}) = \frac{1}{2} \sqrt{\frac{5}{4\pi}} (3\cos^2\Theta_{ij}-1)$.
The angle $\Theta_{ij}$ is the angle $\hat{r}_{ij}$ makes relative to the gradient direction $\hat{g}$, $\cos\Theta_{ij} = \hat{r}_{ij}\cdot\hat{g}$. The expression for $S_{ij}^1$ reduces to
\begin{equation}
    S_{ij}^1 = \frac{2^4c_0}{3D} \int_0^\infty dk \ \frac{h^2(ak)}{k^6} \ \left[j_0(r_{ij}k) - j_2(r_{ij}k) (3\cos^2\Theta_{ij}-1) \right]
\end{equation}
We can make the integral dimensionless by making the variable substitutions $u \equiv ak$ and $n_{ij} \equiv r_{ij}/a$.
\begin{equation}
    S_{ij}^1 = \frac{2^4c_0a^5}{3D} \int_0^\infty du \ \frac{h^2(u)}{u^6} \left[j_0\left(n_{ij}u\right) - j_2\left(n_{ij}u\right) (3\cos^2\Theta_{ij}-1) \right] \label{eq:s1u1}
\end{equation}
We can break up Eq.\ \ref{eq:s1u1} into two integrals and evaluate them individually. Note that the exact solution to either integral depends parametrically on $n_{ij}$ and that $n_{ij}$ is the number of cells radii separating two cells. If we are evaluating the cross-correlations in one cell then $i=j$ and $n_{ii}=0$; on the other hand, if $i\neq j$ then $n_{ij} \geq 2$ in order to eliminate the possibility of overlapping cells. In either case the expression simplifies to:
\begin{equation}
    S_{ij}^1 =
    \begin{cases}
        \frac{4\pi c_0a^5}{45D} &\ i=j \\
        -\frac{\pi c_0a^5}{18D} \frac{1}{n_{ij}^3}(3\cos^2\Theta_{ij}-1) &\ i \neq j, \ n_{ij} \geq 2
    \end{cases} \ .
\end{equation}
Doing the same set of expansions for $S_{ij}^2$ in Eq.\ \ref{eq:Sij2MW}, and performing the same kind of analysis reveals that the gradient depedendent term is asymmetric under exchange of $i$ and $j$. Therefore when calculating the cluster polarization variance all the $S_{ij}^2$ terms will cancel. The variance contributions $V$ and $C$ are
\begin{align}
    V_\text{IC} &= \sum_{i=1}^N \frac{1}{T} S_{ii,z}(0) = \frac{4\pi a^5c_0}{45DT} N \ , \label{eq:ICV1} \\
    C_\text{IC} &= \sum_{i\neq j}^N \frac{1}{T} S_{ij,z}(0) = -\frac{\pi a^5c_0}{18DT} \sum_{i\neq j}^N \frac{(3\cos^2\Theta_{ij}-1)}{n_{ij}^3} \ \label{eq:ICC1},
\end{align}
resulting in the IC collective total variance
\begin{equation} \label{eq:VIC}
    \text{Var}[P_z] = \frac{\pi a^5c_0}{9DT} \left[ \frac{4}{5}N - \frac{1}{2}\sum_{i\neq j}^N \frac{(3\cos^2\Theta_{ij}-1)}{n_{ij}^3} \right] \ ,
\end{equation}
as in Eqs.\ 6 and 7 in the main text.
Next we will show how Eq.\ \ref{eq:VIC} scales for collectives in one, two and three dimensional configurations.

\subsubsection{One Dimensional Chain}

For a one-dimensional chain of IC cells each cell is aligned parallel to the gradient and the angular dependence of $C_\text{IC}$ (Eq.\ \ref{eq:ICC1}) vanishes,
\begin{equation}
    C_\text{IC} = -\frac{\pi a^5c_0}{18DT} \sum_{i\neq j}^N \frac{2}{n_{ij}^3} \ .
\end{equation}
We evaluate the sum:
\begin{equation*}
    \sum_{i\neq j}^N \frac{1}{n_{ij}^3} = 2 \sum_{i<j}^N \frac{1}{n_{ij}^3} = 2 \sum_{i=1}^{N-1} \frac{N-i}{(2i)^3} = \frac{1}{4}(N H_{N-1}^{(3)} - H_{N-1}^{(2)}) \ ,
\end{equation*}
with $H_n^{(m)} = \sum_{k=1}^n \frac{1}{k^m}$ the generalized harmonic number. This results in a total variance of the form
\begin{equation}
    \text{Var}[P_z] = \frac{\pi a^5c_c}{9DT} \left[ \frac{4}{5}N - \frac{1}{8} \left( N H_{N-1}^{(3)} - H_{N-1}^{(2)} \right) \right] \ .
\end{equation}
For large $N$, $H_{N-1}^{(i)}$ approaches a constant for $i \geq 2$. Therefore, we see that $\text{Var}[P_z]$ scales with $N$ for 1D IC collectives as in Table I of the main text.

\subsubsection{Two Dimensional Sheet}

For a two-dimensional sheet of IC cells, pairs of cells can now make a variety of angles with the gradient, and the angular dependence of $C_\text{IC}$ cannot be easily simplified. In order to find the $N$ scaling for $C_\text{IC}$ we calculate the sum numerically. Since the covariances rapidly fall-off as $1/n^3_{ij}$, we only track nearest neighbor pairs that are less than 3 cell radii apart. The resulting numerical solution to the sum in $C_\text{IC}$ is
\begin{equation*}
    \sum_{i\neq j}^N \frac{3\cos^2\Theta_{ij}-1}{n_{ij}^3}
    = 2 \sum_{i<j}^N \frac{3\cos^2\Theta_{ij}-1}{n_{ij}^3}
    = \frac{1}{4} (1.70 N - 2.67 \sqrt{N} + 0.89) \ .
\end{equation*}
Therefore the expression for $C_\text{IC}$ (Eq.\ \ref{eq:ICC1}) simplifies to
\begin{equation}
    C_\text{IC} = -\frac{\pi a^5c_0}{18DT} (0.43 N - 0.67 \sqrt{N} + 0.22) \ .
\end{equation}
The covariance contribution, $C_\text{IC}$, to leading order scales linearly with $N$.
The total variance becomes
\begin{equation}
    \text{Var}[P_z] = \frac{\pi a^5c_c}{9DT} \left( 0.59 N + 0.33 \sqrt{N} - 0.11 \right) \ .
\end{equation}
We see that for large $N$, $\text{Var}[P_z]$ scales with $N$ for 2D IC collectives as in Table I in the main text.

\subsubsection{Three Dimensional Cluster}

To obtain a scaling for $C_\text{IC}$ in a three dimensional cluster we assume that cluster is large, such that $a \ll R$ and $N \gg 1$. For a given cell we can calculate its contribution to $C_\text{IC}$ by considering the covariance contribution it makes with a set of cells a fixed distance away from it. The equidistant cells form a spherical shell with the principal cell in the center. Adapting Eq.\ \ref{eq:ICC1} for a cell and its spherical shell of covariance pairs yields:
\begin{equation}
    C_\text{cell} = -\frac{\pi a^5c_0}{18DT} \frac{1}{n_\text{shell}^3} \sum_{i_\text{shell}} 3\cos^2\Theta_{i}-1 \ ,
\end{equation}
with $n_\text{shell}$ the radius of the shell in terms of cell radii.
Going to continuum we can calculate the contribution from the cell and all its pairs
\begin{equation}
    C_\text{cell} = -\frac{\pi a^5c_0}{18DT n_\text{shell}^3} \int_0^{2\pi} d\phi \int_0^{\pi} d\theta \ \sin\theta (3\cos^2\theta-1) = -\frac{\pi^2 a^5c_0}{9DT n_\text{shell}^3} \int_0^\pi d\theta \ (3\cos^2\sin\theta - \sin\theta) = 0 \ .
\end{equation}
In the last step, we see that the integral vanishes. Thus, the contribution from a single cell and its shell of pairs sum to zero. Repeating this argument for all cells in the cluster results in the total $C_\text{IC} = 0$. Therefore for 3D clusters there is no covariance contribution to the total variance, and $\text{Var}[P_z] = V_\text{IC} \sim N$ as in Table I of the main text.

\subsection{Emergent Chemotaxis Clusters}
For EC the variance in $P_z$ is
\begin{equation}
    \text{Var}[P_z] = \sum_{i=1}^N \text{Var}[p_{i,z}] + \sum_{i\neq j} \text{Cov}[p_{i,z},p_{j,z}] \equiv V_\text{EC} + C_\text{EC} \ .
\end{equation}
The Fourier-transformed fluctuations in IC cell polarization is
\begin{equation}
    \delta\tilde{p}_{i,z}(\vec{k},\omega) = \cos\Theta_i \int_V d^3x \int \frac{d^3k}{(2\pi)^3}
    \delta\tilde{c}(\vec{k},\omega)
    e^{-i\vec{k}\cdot(\vec{x_i}+\vec{x})} \ ,
\end{equation}
with $\Theta_i$ the angle cell $i$ makes with the gradient.
The cross-spectrum for the $z$-component between two cells is
\begin{equation} \label{eq:pijEC1}
\begin{split}
    \langle\delta\tilde{p}_i^*(\vec{k}',\omega')\delta\tilde{p}_j(\vec{k},\omega) \rangle = \cos\Theta_i \cos\Theta_j \int_V d^3x d^3x' \int \frac{d^3kd^3k'}{(2\pi)^6}
    \langle \delta\tilde{c}^*(\vec{k}',\omega')\delta\tilde{c}(\vec{k},\omega) \rangle
    e^{-i\vec{k}\cdot(\vec{x}_j+\vec{x})} e^{i\vec{k}\cdot(\vec{x}_i+\vec{x}')}
\end{split}
\end{equation}
Following the same procedure as in the case of IC, we get an expression for $S_{ij}^1$ for EC:
\begin{equation} \label{eq:SijEC1}
    S_{ij}^1 =
    \begin{cases}
        \frac{16\pi c_0a^5}{15D} \cos^2\Theta_i &\ i=j \\
        \frac{8\pi c_0a^5}{9D} \frac{1}{n_{ij}} \cos\Theta_i \cos\Theta_j &\ i \neq j, \ n_{ij} \geq 2
    \end{cases} \ .
\end{equation}
Since again $S^2_{ij}=0$ by symmetry, the variance for any configuration of EC cells is
\begin{align}
    V_\text{EC} &= \frac{16\pi a^5c_0}{15DT} \sum_{i=1}^{N_\text{edge}} \cos^2\Theta_i \ , \label{eq:ECV1} \\
    C_\text{EC} &= \frac{8\pi a^5c_0}{9DT} \sum_{i\neq j} \frac{\cos\Theta_i \cos\Theta_j}{n_{ij}} \ , \label{eq:ECC1}
\end{align}
as in Eqs.\ 10 and 11 in the main text. The resulting total variance is
\begin{equation} \label{eq:ECvarP}
    \text{Var}[P_z] = \frac{8\pi a^5c_0}{3DT} \left[ \frac{2}{5} \sum_{i=1}^{N_\text{edge}} \cos^2\Theta_i +  \frac{1}{3} \sum_{i\neq j} \frac{\cos\Theta_i \cos\Theta_j}{n_{ij}} \right] \ .
\end{equation}

\subsubsection{One Dimensional Chain}

For a one-dimensional chain of cells only the two cells on the opposing ends are polarized. The cell variance contribution to the total variance therefore does not change with increasing cluster size,
\begin{equation}
    V_\text{EC} = \frac{16\pi a^5c_0}{15DT} \sum_{i=1}^{N_\text{edge}} \cos^2\Theta_i = \frac{32\pi a^5c_0}{15DT} \ .
\end{equation}
Therefore $V_\text{EC} \sim N^0$ for 1D collectives. For $C_\text{EC}$ the distance between the two edge cells increases by two cell radii for each cell added to the chain:
\begin{equation}
    C_\text{EC} = \frac{8\pi a^5c_0}{9DT} \sum_{i\neq j} \frac{\cos\Theta_i \cos\Theta_j}{n_{ij}} = -\frac{8\pi a^5c_0}{9DT} \frac{1}{2(N-1)} \ .
\end{equation}
So $C_\text{EC} \sim N^{-1}$ for 1D collectives. To leading order in $N$ the total collective variance depends only on $V_\text{EC}$:
\begin{equation}
    \text{Var}[P_z] = \frac{32\pi a^5c_0}{15DT} \ ,
\end{equation}
and so $\text{Var}[P_z]$ does not depend on collective size for 1D EC as in Table I of the main text.

\subsubsection{Two Dimensional Sheet}

In order to evaluate the variance for a two-dimensional disc of cells we will approximate the sums as integrals over the circumference of the disc as we did in evaluating the mean polarization. Assuming that $a \ll R$ Eq.\ \ref{eq:ECV1} can be written as an integral
\begin{equation}
    V_\text{EC} = \frac{16\pi a^5c_0}{15DT} \frac{R}{2a} \int_0^{2\pi} d\theta \ \cos^2\theta \ .
\end{equation}
Using the relation $N=(R/a)^2$ yields
\begin{equation}
    V_\text{EC} = \frac{8\pi^2 a^5c_0}{15DT} \sqrt{N} \ .
\end{equation}
Hence for 2D EC, the variance contribution $V_\text{EC}$ scales as $\sqrt{N}$.
In order to determine how $C_\text{EC}$ scales with $N$ we approximate the sums over $i$ and $j$ as a double integral, again assuming that $a \ll R$.
\begin{equation}
    C_\text{EC} = \frac{16\pi a^5c_0}{9DT}
    \left( \frac{R}{2a^2} \right) \int_{\Delta/2}^{2\pi-\Delta/2} d\theta_1 \int_{\theta_1+\Delta/2}^{2\pi} d\theta_2 \ \frac{\cos\theta_1 \cos\theta_2}{n(\theta_1,\theta_2)}
\end{equation}
Here $\Delta = 2a/R$ is the anguler separation between two edge cells, and
\begin{equation*}
    n(\theta_1,\theta_2) = \frac{2R}{a} \sin \left( \frac{1}{2}(\theta_2-\theta_1) \right)
\end{equation*}
is the number of cell radii separating two edge cells. Using this expression for $n(\theta_1,\theta_2)$ we evaluate the integral over $\theta_2$:
\begin{equation}
    \begin{split}
        \left( \frac{R}{2a^2} \right) \int_{\Delta/2}^{2\pi-\Delta/2} &d\theta_1 \int_{\theta_1+\Delta/2}^{2\pi} d\theta_2 \ \frac{\cos\theta_1 \cos\theta_2}{n(\theta_1,\theta_2)} \\
        &= \frac{R}{8a} \int_{\Delta/2}^{2\pi-\Delta/2} d\theta_1 \cos\theta_1 [ -4\left( \cos(\theta_1/2) + \cos(\theta_1+\Delta/2) \right)
        -2 \cos\theta_1 \log\left( \tan(\Delta/4) \tan(\theta_1/4) \right) ]
    \end{split}
\end{equation}
Breaking up the integral into four separate terms we find:
\begin{align*}
    &\int_{\Delta/2}^{2\pi-\Delta/2} d\theta_1 \cos\theta_1 \cos(\theta_1/2) = 0 \ , \\
    &\int_{\Delta/2}^{2\pi-\Delta/2} d\theta_1 \cos\theta_1 \cos(\theta_1+\Delta/2) = -\frac{1}{2}\cos(\Delta/2) (\Delta+\sin\Delta-2\pi) \ , \\
    &\int_{\Delta/2}^{2\pi-\Delta/2} d\theta_1 \cos\theta_1 \log\left( \tan(\Delta/4)\right) = -\frac{1}{2} (\Delta+\sin\Delta-2\pi) \tan(\Delta/4) \ , \\
    &\int_{\Delta/2}^{2\pi-\Delta/2} d\theta_1 \cos\theta_1 \log\left( \tan(\theta_1/4) \right) = 0 \ .
\end{align*}
The first and last integrals are equal to zero since the integrands are odd functions over the range $[0,2\pi]$. With these results, the whole expression simplifies to
\begin{equation}
    C_\text{EC} = \frac{16\pi a^5c_0}{9DT} \frac{1}{4}\sqrt{N} \left( \frac{1}{2}\log N + \log 2 -2 \right) \left( \pi - \frac{2}{\sqrt{N}} \right)
\end{equation}
Keeping only the leading order terms in $N$ yields
\begin{equation}
    C_\text{EC} = \frac{2\pi a^5c_0}{9DT} \sqrt{N} \log N \ .
\end{equation}
The resulting total variance is
\begin{equation}
    \text{Var}[P_z] = \frac{8\pi a^5c_0}{3DT} \sqrt{N} \left[ \frac{\pi}{5} +  \frac{1}{12} \log N \right] \ ,
\end{equation}
which to to leading order scales as $\sqrt{N} \log N$ as in Table I of the main text.

\subsubsection{Three Dimensional Cluster}

For the three-dimensional cluster, numerical methods must be used in order to find the scaling properties of the variance. We numerically evaluate the total variance (Eq.\ \ref{eq:ECvarP}) on a cubic lattice and obtain the following results.

\begin{figure}[ht]
    \centering
        \includegraphics[width=0.50\textwidth]{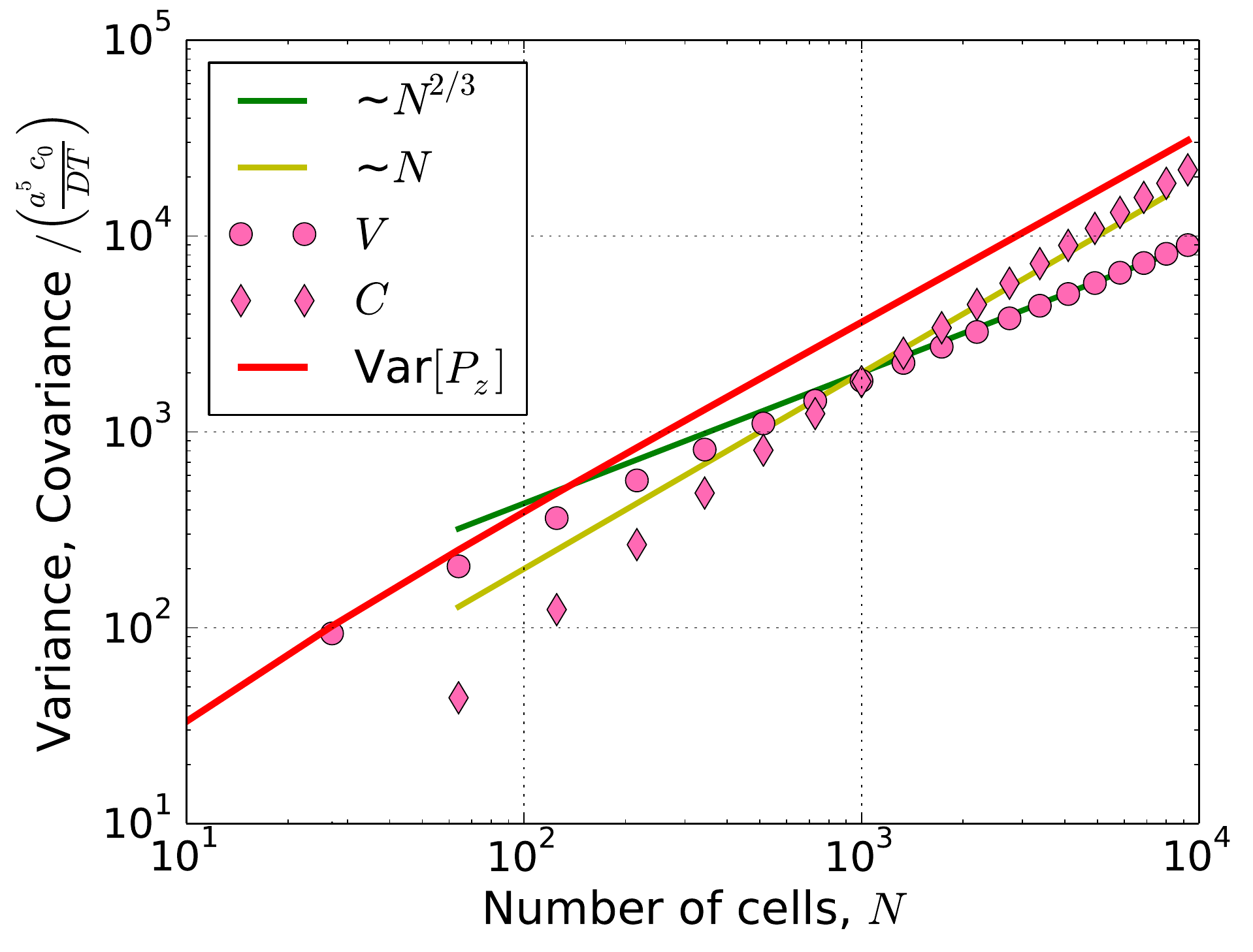}
    \caption{$\text{Var}[P_z]$ for a 3D cluster of EC cells. Cluster variance shown in red. Pink circles are the single cell variance contributions $V$, and pink diamonds are the cell-cell covariance contributions $C$.} \label{fig:S2}
\end{figure}

The numerical results [Fig.\ \ref{fig:S2}] show that $V \sim N^{2/3}$ since the number of edge cells also scales as $N^{2/3}$. We also find that $C \sim N$; the covariance contribution to the total cluster polarization grows linearly with $N$. For large clusters the $N$ scaling dominates the behavior of $\text{Var}[P_z]$. Therefore, in 3D the leading order scaling for the variance is
$\text{Var}[P_z] \sim N$ as in Table I of the main text.

\section{Testing Analytic Model Assumptions}

\begin{figure}[ht]
    \centering
        \includegraphics[width=0.65\textwidth]{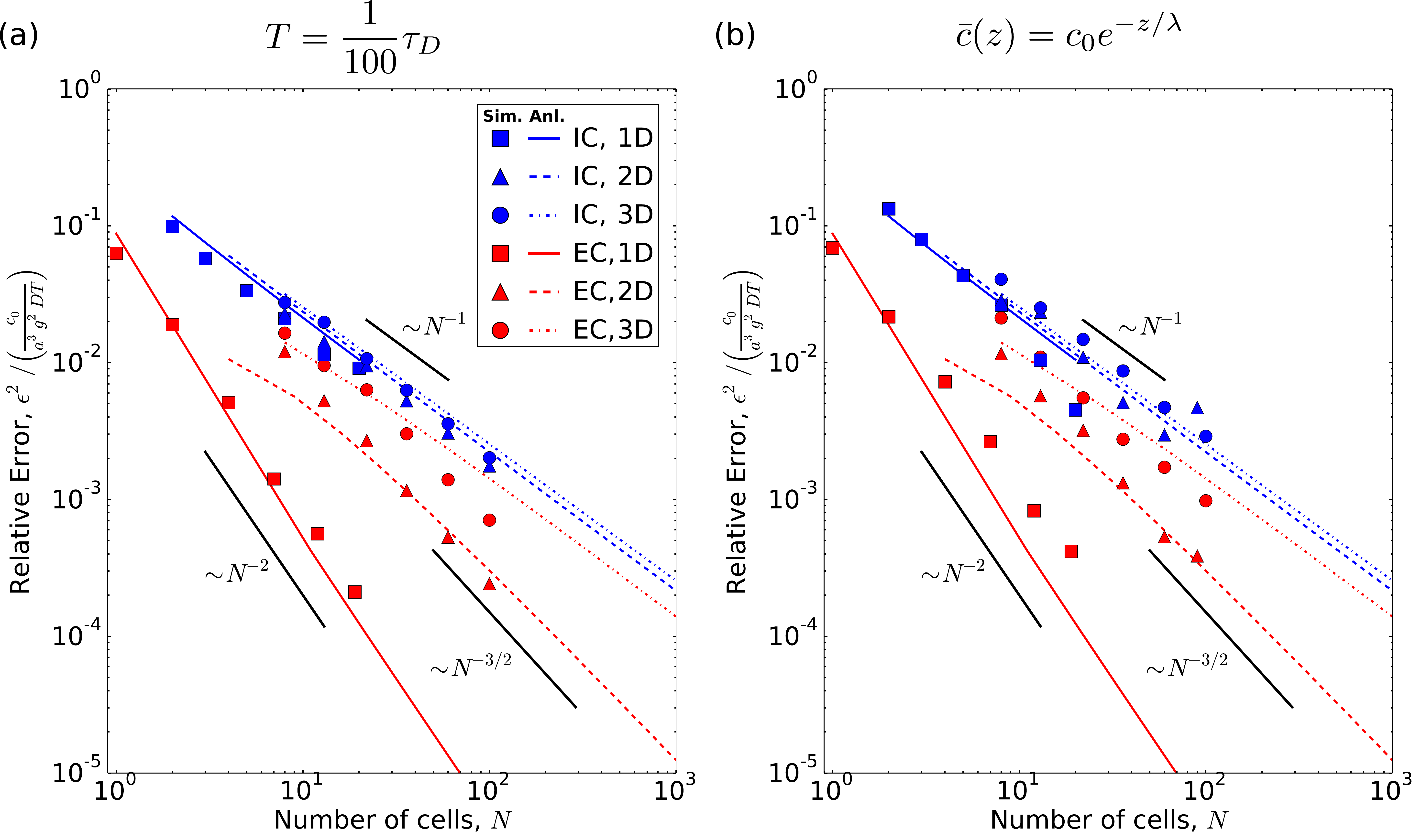}
    \caption{(a) Short-time integration relative error results. Data points are of simulations for $T= \frac{1}{100} \tau_D$. (b) Exponential concentration profile relative error results. The mean concentration profile is $\bar{c}(z) = c_0 e^{-z/\lambda}$, the lengthscale $\lambda=\sqrt{D/\beta}$ is set by the diffusion coefficient $D$ and the molecule decay rate $\beta$. Lines are from original analytical predictions using $T > \tau_D$ and a linear concentration profile.} \label{fig:S1}
\end{figure}

Simulations are performed to test model behavior when assumptions used to derive analytic results are relaxed. In Fig.\ \ref{fig:S1}(a) we relax the assumption that the integration time $T$ must be larger than the timescale for diffusion $\tau_D \sim R^2/D$. We find that $\epsilon^2$ scales the same way as previously predicted for both EC and IC, even when $T = \tau_D/100$. The only exception is that $\epsilon^2$ for 3D EC [Fig.\ \ref{fig:S1}(a), red circles] has a more negative power-law dependence on $N$ than the expected $\sim N^{-1}$. The shorter integration time results in decreased correlations between edge cells which when $T>\tau_D$ results in $C\sim N$. Hence with
$T < \tau_D$
the total variance is less dependent on $C$, and $V \sim N^{2/3}$ becomes the dominant contribution to $\text{Var}[P]$. This results in a steeper scaling of $\epsilon^2$ closer to
$\text{Var}[P]/\langle P \rangle^2 \sim N^{2/3}/N^2 = N^{-4/3}$.
Interestingly, we see that relaxing the assumption $T \gg \tau_D$ results in improved precision for EC over IC not just in 1D and 2D but also in 3D configurations.

In Fig.\ \ref{fig:S1}(b) we change the concentration profile from linear to exponential which has a mean concentration of
$\bar{c}(z) = c_0 e^{-z/\lambda}$.
The lengthscale $\lambda=\sqrt{D/\beta}$ depends on the diffusion coefficient and the molecule degradation rate $\beta$. In Fig.\ \ref{fig:S1}(b) the simulation results are for $\lambda > a$. We find that $\epsilon^2$ is in very good agreement with our original analytic predictions. The only exception is that due to the exponential profile, $\langle P \rangle$ for 1D EC (Fig.\ \ref{fig:S1}(b), red squares) is non-linear in $N$ causing the relative error data to scale less steeply than the expected $N^{-2}$.

\section{Error Estimates in Biological Systems}
In this section we discuss how IC and EC relative error values for experimental systems can be estimated. The relative errors are given for mammary epithelial organoids (Ref.\ [5,31] of the main text) and collectives of malignant lymphocytes (Ref.\ [13] of the main text). We also discuss how the errors in these systems may be measured using the chemotactic index.

The numerical values for the error
$\epsilon$
can be calculated using our IC and EC analytical predictions as well as values for the biological parameters
$a, c_0, g, D$, and $T$. We can express the relative error as
$\epsilon^2 = \alpha \ \frac{c_0}{a^3 g^2 D T}$ with $\alpha$ being the numerical factor dependent on the model choice (IC or EC), the dimensionality of the collective (1D, 2D or 3D), and the collective size. We refer to $\alpha$ as the scaled relative error and it is equivalent to the results plotted in Fig.\ 2. Therefore, for a given set of biological parameters, we can calculate $\epsilon$ for a given collective migration mechanism, dimensionality, and size.

Refs.\ [5,31] study the sensory capabilities of epithelial organoids, and from these studies we obtain the following biological parameters:
$a = 10\mu\text{m}$, $c_0 = 10\text{nM}$, $g = 0.5 \text{nM/mm}$,
$D \approx 50 \mu\text{m}^2/\text{s}$, and $T \approx 10 \text{s}$.
The integration time was not directly measured, and was taken from their theoretical arguments (Ref.\ [31]).
The biological parameters result in $c_0 / (a^3g^2DT) \approx 13.3$.
Specifically, the sensory unit of the organoid likely only comprises $N \sim 5$ cells within a small one-dimensional, finger-like protrusion from the organoid (Ref.\ [5]). Combining these results we find errors of
$\epsilon_\text{IC} \approx 73\%$, and $\epsilon_\text{EC} \approx 16\%$.
Ref.\ [13] studies chemotaxis of malignant lymphocyte collectives, from which we obtain the following biological parameters:
$a = 10\mu\text{m}$,
$c_0 \approx 12.5 \text{ng/mL}$, $g \approx 2.5\cdot 10^{-2} \text{ng/mL}$,
$D \approx 100 \mu\text{m}^2/\text{s}$, and $T \approx 10 \text{s}$.
We estimate the integration time by observing from that study that lymphocytes generally migrate a cell length on the order of 1 minute, requiring an integration time that is shorter than that. Converting the chemical concentrations to molar using the molecular weight of the attractant, $10\text{kDa}$, these values result in
$c_0 / (a^3g^2DT) \approx 0.33$.
Chemotaxis is reported to occur as 2D collectives with an average size of $N \sim 15$ cells, resulting in errors of
$\epsilon_\text{IC} \approx 8.1\%$, and $\epsilon_\text{EC} \approx 3.6\%$.
These two examples illustrate that the differences between IC and EC models of collective migration can be substantial. In the case of the epithelial organoids both models are capable of collective migration that is not drowned out by noise, but the choice of EC provides a fourfold reduction in error. Similarly, malignant lymphocyte collectives gain a twofold reduction in error if using EC.


\begin{thebibliography}{39}
\bibitem[{foo({\natexlab{b}})}]{footnote2}
\bibinfo{note}{Treating cells as permeable spheres neglects receptor binding.
  We expect that the dimensionality-dependent scaling results derived herein
  will not change for reversible receptor binding but may change if binding is
  irreversible, as irreversible binding fundamentally changes the boundary
  conditions that cells present to the molecule field.}
\bibitem[{foo({\natexlab{c}})}]{footnote3}
\bibinfo{note}{Although cells are soft, here we assume as in previous studies
  \cite{szabo2006phase,camley2016emergent,camley2016collective} that their
  polarization vectors may be summed as if the cells were rigid bodies.}
\bibitem[{foo({\natexlab{d}})}]{footnote4}
\bibinfo{note}{Although we only study idealized collective shapes, the
  dimensionality-dependent scalings we derive are likely to persist in these
  specific cellular systems because the scalings are independent of the exact shapes used.}
\end{thebibliography}
\end{document}